\documentclass{PoS}
\input{epsf.tex}
\usepackage{amssymb,epsfig}
\PoS{PoS(LAT2005)249}

\title{Fermion propagators in QED$_3$ with velocity anisotropies }

\ShortTitle{Fermion propagators in QED$_3$ with velocity anisotropies }

\author{\speaker{Iorwerth Owain Thomas}\\
        University of Wales Swansea\\
        E-mail: \email{pyiori@swansea.ac.uk}}

\author{Simon Hands\\
        University of Wales Swansea\\
        E-mail: \email{s.hands@swansea.ac.uk}}

\abstract{QED$_3$ with fermi and gap anisotropies is considered to be a candidate effective field theory of high temperature superconductors.  Simulations of a variant of the theory have demonstrated that there is evidence consistent with a phase transition to a chirally restored phase as the velocity anisotropy increases, and that the correlation lengths of pions in spatial directions behave in a manner consistent with the anisotropy being a relevant parameter of the theory. We present the first measurements of the fermion propagator in Landau gauge for this theory and suggest that the structure of this theory is more complex than previously thought.
}

\FullConference{XXIIIrd International Symposium on Lattice Field Theory\\
		 25-30 July 2005\\
		 Trinity College, Dublin, Ireland}

\begin{document}

\section{Introduction}

QED$_3$ has been the subject of recent interest in condensed matter theory as a candidate effective theory of the underdoped phases of cuprate high-temperature superconductors \cite{kn:Tes0,kn:Herb0}.  

These effective theories have an interesting peculiarity: since we have no reason to suspect that Lorentz invariance is respected in a condensed matter system, we can introduce `velocities' $v_F$ and $v_\Delta$ that parametrise the deformation of the Fermi surface of the underlying condensed matter theory.  At low energies, our action is \cite{kn:Herb1}:
\begin{eqnarray}
	S= \int d^{2}{\bf r} \, d\tau \, \{\bar{\Psi}_{1}[\gamma_{0}(\partial_{\tau} +ia_{\tau}) + \frac{\delta}{\sqrt{\kappa}}\gamma_{1}(\partial_{x} +ia_{x})
&+& \delta\sqrt{\kappa}\gamma_{2}(\partial_{y} +ia_{y})]\Psi_{1}\nonumber\\
&+& (1 \rightarrow 2, x \leftrightarrow y) 
+ \frac{1}{4e^{2}}F_{\mu\nu}F_{\mu\nu}\} \label{eqn:fullQED3}
\end{eqnarray}
\noindent where $\Psi_i$ is an excitation around one of the two pairs of nodes possessed by the d-wave Fermi surface, $\kappa=\frac{v_F}{v_\Delta}$ (this quantity is proportional to the doping of a cuprate sample), $\delta=\sqrt{v_\Delta v_F}$ and the gauge fields model statistical interactions between excitations at the nodes.

In this approach, it has been suggested that the chirally-broken phase of the model corresponds to the onset of a `spin-density wave' phase \cite{kn:Herb0,kn:Tes1}, and that the chirally symmetric phase to the so-called `pseudogap' phase, some of whose peculiar properties, it is hoped, may be explained by chirally-symmetric QED$_3$ \cite{kn:Tes0}.

In order that we might make use of an exact HMC algorithm, we use the following simplified version of (\ref{eqn:fullQED3}), which we expect to capture most of the physics of the system:
\begin{equation}
S =\sum_{i=1}^{2} \sum_{x,x'} a^3\bar{\chi_{i}}(x) M_{x,x'} \chi_{i}(x') 
%	\nonumber\\
+\frac{\beta}{2} \!\!\sum_{x,\mu<\nu}\!\! a^3\Theta_{\mu\nu}^{2}(x) 
\label{eqn:lattact}
\end{equation}
\noindent where $M_{x,x'}$ is the staggered fermion matrix, modified as follows,
\begin{equation}
	M_{x,x'} =  {1\over2a}\sum_{\mu=1}^{3} \xi_{\mu}(x) 
[\delta_{x',x+\hat{\mu}} U_{x\mu} -
\delta_{x',x-\hat\mu}U_{x'\mu}^\dagger]
%\nonumber\\ 
+ m\delta_{\mu\nu},  \label{eqn:fermion_matrix}
\end{equation}
\noindent where $\xi_{\mu}(x) = \lambda_{\mu} \eta_{\mu}(x)$:  $\eta_{\mu}(x)=(-1)^{x_{1} + ... + x_{\mu-1}}$, which is the Kawamoto-Smit phase factor ( where $x_1=x$, $x_2=y$ and 
$x_3=t$), and $\lambda_{x}=\kappa^{- \frac{1}{2}}$, $\lambda_{y}=\kappa^{\frac{1}{2}}$, $\lambda_{t}=1$. 

$U_{x\mu}\equiv \exp(ia\theta_{x\mu})$ is the gauge parallel transporter, and the gauge action has the {\em non-compact} form 
\begin{equation}
	\Theta_{\mu\nu}(x) = {1\over a}[\Delta_{\mu}^{+}\theta_{\nu}(x)-  
\Delta_{\nu}^{+}\theta_{\mu}(x)]. 
\end{equation}
In \cite{Hands:2004ex}, we presented evidence from simulations on a $16^3$ lattice of a phase transition to what appears to be a chirally symmetric state as $\kappa$ increases above a critical value, $\kappa_c\approx 4.35$, and that, contrary to previous analytical studies, the renormalised value of $\kappa$ (when defined as the ratio between the pion screening lengths in the $x$- and $y$-directions) is a relevant parameter (i.e. $(\kappa_{\pi ren} -1)>(\kappa_{bare}-1)$).  In what follows, we present measurements of the fermion propagator in Landau gauge, whose behaviour is thought (not uncontroversially) \cite{kn:Tes0,Khveshchenko:2001jn}, to provide some explanation for the behaviour of the pseudogap phase.

\section{Gauge Fixing}

The fermion propagator is not a gauge invariant quantity; prior to measuring it we must numerically impose a gauge fixing condition.  We do so as follows \cite{Nakamura:1990dq,Gockeler:1990bc}:

\begin{itemize} 
\item We impose `minimal' Landau gauge, corresponding to the continuum condition $\partial_\mu\theta_\mu=0$:
\begin{equation}
\sum_\mu \theta_{\mu\,x} - \theta_{\mu\,x-\hat{\mu}}=0.
\end{equation}

This corresponds to the minimum of a functional $F[\theta]$.  For reasons of practicality, in numerical simulations one imposes the condition up to a residual value, $R$, which we choose to be the floating point value $10^{-6}$.

\item To remove all remaining gauge degrees of freedom (largely due to zero modes that are an artefact of the finite volume of the lattice), we rotate $\theta_{\mu\,x}\mapsto \theta_{\mu\,x} \pm n \frac{2\pi}{L_{\mu}}$, $\, \forall \theta_{\mu\,x}$,$\, \mu$, where $n$ is an integer and $L_\mu$ is the size of the lattice in that direction, such that:
\begin{equation}
-\frac{2\pi}{L_\mu}<\bar{\theta}_\mu\leq\frac{2\pi}{L_\mu}, 
\end{equation}
\noindent where $\bar{\theta}_\mu=\frac{1}{V}\sum_x \theta_{\mu\,x}$ .
\end{itemize}

We call this the 'modified non-compact iterative Landau gauge' or `miLandau gauge' (by analogy with the similarly named condition proposed in \cite{Durr:2002jc} for compact QED).

It is, of course, necessary to check that there are no false minima of $F[\theta]$ (Gribov copies) that might contaminate our measurements; while this gauge condition has been used before, this check has yet to be carried out for noncompact QED$_3$, simply as it seems unlikely that there could be any such problem.  It would be good to confirm such an intuition (particularly as compact QED$_3$ has a non-trivial Gribov problem \cite{Mitrjushkin:1996fw}).

In order to do this, we followed the method of \cite{Bogolubsky:1999cb}:  From a set of mother configurations, we produced 3 sets of 500 daughter configurations via the following gauge transformations:

\begin{itemize}
\item[{\bf A:}] $\theta_{\mu\,x}\mapsto \theta_{\mu\,x} -\alpha_{x+\hat{\mu}}+\alpha_x$, where $\alpha_x$ is a random number.
\item[{\bf B:}] $\theta_{\mu\,x}\mapsto \theta_{\mu\,x} \pm n \frac{2\pi}{L_{\mu}}, \, \forall \theta_{\mu\,x}$ in the direction $\mu$, where $n$ is randomly chosen to be 0, 1, or -1.
\item[{\bf C:}] We perform transformation {\bf A} and then {\bf B}.
\end{itemize}

\begin{figure}[htb]
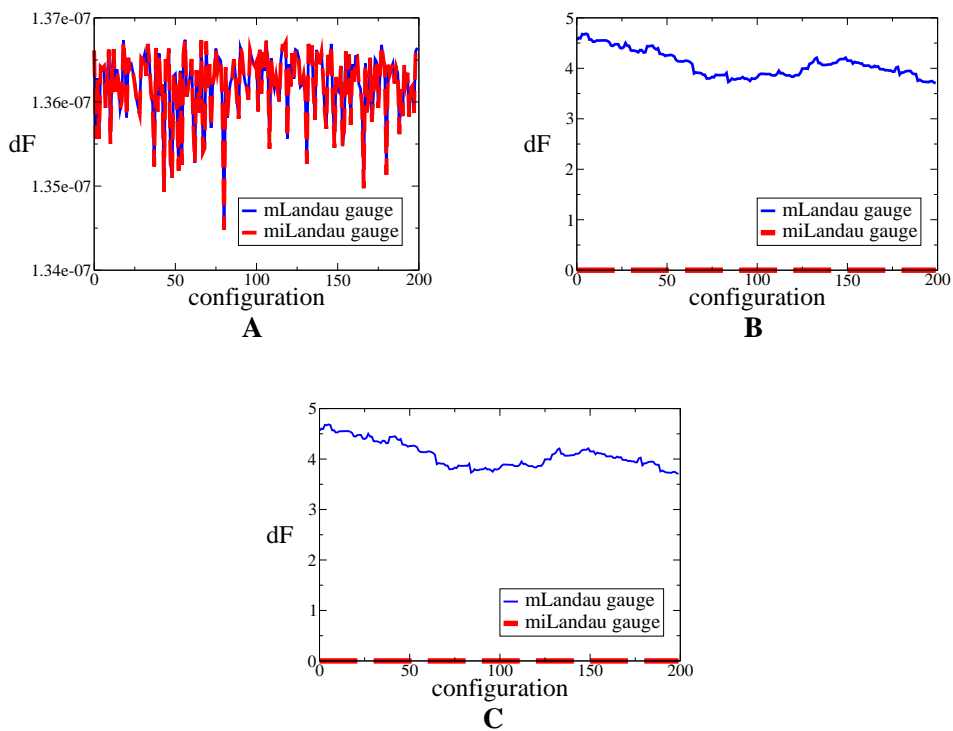

\vspace{0cm}
\begin{center}
\epsfig{file=Graphs/norotvar.eps, height=4cm}
\hspace{1cm}
\epsfig{file=Graphs/onlyvar.eps,height=4cm}
\end{center}
\vspace{-.6cm}
\hspace{4.4cm}{\bf A}\hspace{6.4cm}{\bf B}
\vspace{.5cm}
\begin{center}
\epsfig{file=Graphs/rotvar.eps,height=4cm}
\end{center}
\vspace{-.5cm}
\hspace{7.6cm}{\bf C}
\caption{\small Variances for the sets of gauge transformations {\bf A}, {\bf B} and {\bf C}.}
\vspace{-.3cm}
\label{fig:Grib}
\end{figure}

We then define a `variance' $dF=\max_{ij}\left[F[\theta]_i - F[\theta]_j\right]$, where $i$, $j$ run over a mother configuration and all its daughters.  If, for a given mother and its offspring, $dF<R$, then for all practical purposes, there are no Gribov copies.  Our results are presented in Figure \ref{fig:Grib} for a $16^3$ lattice with $m=0.03$,  $\beta=0.2$ and $\kappa=1.00$ (we obtained similar results for a value of $\kappa=10.00$).  We see that for set {\bf A}, there are no gauge copies for either gauge condition.  However, for {\bf B} and {\bf C}, we see that there are Gribov copies present for mLandau gauge, but not for miLandau gauge.  Furthermore, the variance is more-or-less identical for both of these groups, which implies that the source of Gribov copies are rotations of kind found in group {\bf B}.  We conclude that, in line with our expectations, the miLandau conditions fix the gauge completely.
 
\section{Fermion Propagator}

Fermion propagators for $\beta=0.2$, $m=0.03$ on a $16^3$ lattice were measured over 30,000 trajectories for $1\leq\kappa\leq10$, the extremely large number of trajectories being necessitated by the noisiness of the observable.  A very slight tendency towards  sawtooth behaviour was noted in the time-direction propagators above $\kappa_c$ as the value of the propagator on even-numbered slices began to drop; this was more pronounced in the $y$-direction and absent in the $x$-direction.  It is not clear why this should be; it may have something to do with the restoration of chiral symmetry, though it does not quite replicate the expected behaviour of a fermion propagator under such circumstances (\cite{Hands:2001cs} gives an example of what we expect) .  Ideally, one would like to fit data with this behaviour to a four parameter fit; however, these fits proved to be unstable on this lattice size, and so we fitted a two-parameter curve to the odd timeslices only.  

\begin{figure}[htb]
\vspace{.1cm}
\begin{center}
\epsfig{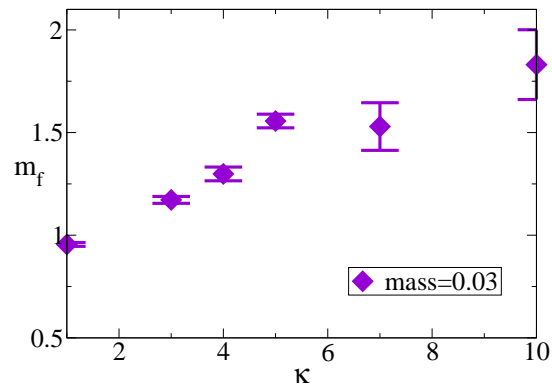}
\end{center}
\vspace{-.7cm}
\caption{\small Fermion masses for various values of $\kappa$.}
\label{fig:fermass}
\vspace{-.3cm}
\end{figure}

\begin{figure}[htb]
\vspace{1.0cm}
\begin{center}
\epsfig{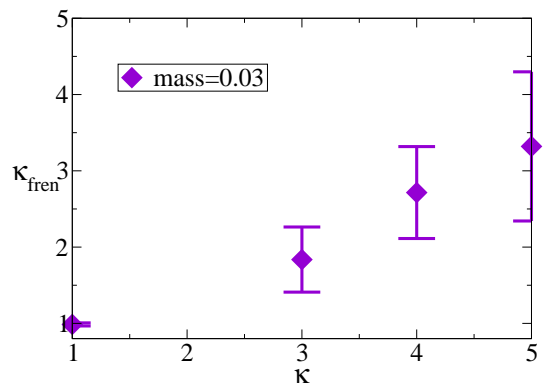}
\end{center}
\vspace{-.7cm}
\caption{\small The renormalised anisotropy, $\kappa_{f ren}$, for various $\kappa$.}
\vspace{-.3cm}
\label{fig:ferkappa}
\end{figure}

Our results are presented in Figures \ref{fig:fermass} and \ref{fig:ferkappa}, where $\kappa_{fr}$, the renormalised anisotropy with respect to fermions, is the ratio of the fermion screening mass in the $x$-direction with that in the $y$-direction.  We may draw two conclusions here: firstly, that fermions retain a dynamically generated mass above $\kappa_c$, and secondly that $\kappa_{fren}-1<\kappa_{bare}-1$ -- that is, $\kappa$ is an {\em irrelevant} parameter with respect to fermions.  This last is quite puzzling, given that $\kappa_{\pi ren}$ is a relevant parameter of the theory, and may take some thought to resolve.

This retention of the dynamically generated mass into an apparently chirally-symmetric phase is more readily explicable, however, and has also been observed in the $2+1$D Gross-Neveu model at finite temperature \cite{Hands:2001cs}.  Witten \cite{Witten:1978qu} suggests that such behaviour may be observed if:

\begin{itemize}
\item The physical fermion corresponds to a branch-cut, and not a pole, and is constantly interacting with massless, non-Goldstone bosons.

\item $\langle \bar{\psi}\psi\rangle$, the chiral condensate, is {\em close} to having long-range order -- for example, it has a phase parameter that slowly varies with spatial/temporal position.  This would entail an overall non-zero magnitude of the order parameter, but fermions would still experience a dynamically generated mass.
\end{itemize}

If correct, this behaviour is unanticipated by any analytic prediction that we are aware of.  Because of this, it renders the status of the condensed matter model a little unclear.

Nevertheless, more research is needed before this can be clarified -- the thermodynamic and continuum limits must be taken, for example, and it would be good to have some measurements of the fermion mass closer to the chiral limit.  Even if this theory proves not to be viable as a model of a condensed matter theory, we hope that its behaviour is interesting enough in its own right to be deserving of study.

\section{Acknowledgements}

SJH is supported by a PPARC Senior Research Fellowship, IOT by a University of Wales Postgraduate Research Scholarship.

\end{document}